\documentclass{aa}

\usepackage{natbib,twoopt}
\usepackage{graphicx} 
\usepackage{xcolor}

\usepackage{amsmath}

\usepackage[breaklinks=true]{hyperref} 
\bibpunct{(}{)}{;}{a}{}{,} 
\makeatletter
\newcommandtwoopt{\citeads}[3][][]{\href{http://adsabs.harvard.edu/abs/#3}%
{\def\hyper@linkstart##1##2{}%
\let\hyper@linkend\@empty\citealp[#1][#2]{#3}}}
\newcommandtwoopt{\citepads}[3][][]{\href{http://adsabs.harvard.edu/abs/#3}%
{\def\hyper@linkstart##1##2{}%
\let\hyper@linkend\@empty\citep[#1][#2]{#3}}}
\newcommandtwoopt{\citetads}[3][][]{\href{http://adsabs.harvard.edu/abs/#3}%
{\def\hyper@linkstart##1##2{}%
\let\hyper@linkend\@empty\citet[#1][#2]{#3}}}
\newcommandtwoopt{\citeyearads}[3][][]%
{\href{http://adsabs.harvard.edu/abs/#3}
{\def\hyper@linkstart##1##2{}%
\let\hyper@linkend\@empty\citeyear[#1][#2]{#3}}}
\makeatother

\usepackage[normalem]{ulem}

\usepackage[varg]{txfonts}

\title{The spatial structure of solar magnetoconvection at  scales up to $60$~Mm.}
\titlerunning{Doppler flow measurements and density depressions at supergranular scales.}
\author{
D. Przybylski \inst{1}
\and
R. H. Cameron \inst{1}
}

\institute{
{Max-Planck-Institut f{\"u}r Sonnensystemforschung, G{\"o}ttingen, Germany}\\
\email{cameron@mps.mpg.de}}

\date{}

\begin{document} 

\abstract
{While convective motions at the scale of granules on the solar surface are well understood, those of mesogranular and supergranulation are not. At still larger scales, the lack of giant cells is a major mystery.}
{We aim to slightly improve understanding of the  radial and horizontal flows associated at mesogranular scales and supergranulation.}
{We perform a deep MURaM small-scale dynamo simulation, extending from $50$~Mm below the solar surface and 
with a horizontal size of 60~Mm and sufficient 
resolution to begin to properly resolve the magnetic bright points in intergranular lanes.
We determine the structure of the convective modes from the simulations at different scales, and compare the power spectra to those which have been previously observed.}
{At all scales which are resolved by the simulation the simulated flows agree with the amplitude implied by  Doppler measurements and local correlation tracking (LCT) measurements. The power in the vertical Doppler velocities at supergranular and larger scales is shown to be influenced by density perturbations near the surface. 
The structure of the modes at depth is simple as expected from the previous study of \cite{LordCameron2014} except the  horizontal forcing scale scales like $H_\rho^{2/3}$ instead of $H\rho$.}
{The simulations reproduce the observations up to scales of $60$~Mm. Within the simulated range, $l>=70$, we do not see the excess power associated with the convective conundrum.}
\maketitle
\nolinenumbers
\section{Introduction}

The structure of solar convection at the scales of granules ($1$ to $2$~Mm) was essentially solved by \cite{SteinNordlund1989}, see also the review of \cite{NordlundStein2009}. Convection at scales greater than that of the granules remains poorly understood \citep{HanasogeDuvall2012,HottaBekki2023, BirchProxauf2024}, although high-resolution simulations including the effects of the small-scale dynamo field has been shown to reduce the discrepancy between the observations and numerical simulations \citep{HottaKusano2021, HottaKusano2022}. 
In this paper we apply and slightly extend the analysis presented in \cite{LordCameron2014} to a simulation at higher resolution and including the magnetic fields generated by a small-scale dynamo.  
We share the same aims as the previous papers on this topic, to improving our understanding of convection at scales larger than granules. 

In this paper we present MURaM simulations of solar magnetoconvection in a box which extends 50~Mm below the solar surface and 60~Mm in the two horizontal directions. We extract the convective modes at different spatial scales from the simulation, and compare the power spectra from the simulations with those based on solar observations.

\section{Methods}
\subsection{MURaM simulation details}
We use the MURaM code \citep{VoglerShelyag2005,Rempel2017} to perform a radiation-magnetohydrodynamic simulation of the upper $50$~Mm of the solar convection zone and photosphere. The code solves the MHD equations in conservative form, and uses a short-characteristics solver with a grey treatment of radiation transport. The equation of state (EoS) is from FreeEoS \citep{Irwin2012} and the opacities are calculated using the MPS-ATLAS code \citep{Witzke2021}, both using the abundances of \citet{Asplund2009}. In order to weaken the time step constraint set by the high Alfv\'en speed in the atmosphere above strong magnetic field concentrations, we use a semi-relativistic treatment (Boris correction) to limit the Alfv\'en speed to $100~\mathrm{km\;s}^{-1}$ \citep{Boris1970,Rempel2017}. The code includes a numerical slope limited diffusion scheme \cite{Rempel2014,Rempel2017}. 

The simulation covers an area of $60\times60~\mathrm{Mm}$ in the horizontal ($x$ and $y$) directions. In the vertical ($z$) direction the simulation spans the upper convection zone and photosphere, from $50~\mathrm{Mm}$ below the average $\tau_{500} = 1$ layer, to $1~\mathrm{Mm}$ above it. The upper boundary is open to outflows, but closed to inflows. The field at the top surface is matched to a potential magnetic field, 
the lower boundary is open and allows advection of the magnetic field \citep[the OsB of][]{Rempel2014}. The horizontal boundaries are periodic.

The simulation is started from a horizontally averaged initial condition, based on a G2V stellar structure model with constant gravity. We start with a low-resolution hydrodynamic simulation, with a grid size of $100$~km in each of the horizontal directions and $40.48$~km in the vertical direction. The simulation is run for $41~\mathrm{days}$ until convection stabilizes throughout the box. A small, zero net-flux seed field is then added, this consists of a random vertical field in $x$ and $y$ and constant in $z$. The simulation is continued to allow the SSD mechanism to enhance the field. It is run for another $21~\mathrm{days}$ until the net horizontally averaged energy flux stabilizes to transport a solar luminosity. We then double the resolution in all directions to $50\times50\times20.24~\mathrm{km}$, which gives $1200\times1200\times 2520$ grid points and  run for another $3~\mathrm{days}$. As a final step, we increase the resolution again in the horizontal directions to $25\times25~\mathrm{km}$ corresponding to $2400\times2400 \times 2520$ grid points. We continue the simulation for another $3$ hours, after which we ran the code for an additional $24$ hours and for which we saved horizontal slices at different depths with an approximately $130$ second (500 time steps) cadence. In Appendix \ref{app:stability} we show the time evolution of the energies and total energy flux in the simulation to demonstrate that it is sufficiently close to the relaxed state.

\subsection{Fourier analysis, and mode structure}

To determine the spatial structure of the convective cells in the simulation we first move into spectral space in the two horizontal directions by taking a discrete Fourier transform, and assume the mode is separable in time and space, so that
\begin{equation}
    q(x,y,z,t)= \sum_{\bf k}\widetilde{q}(z) f(t) {\mathrm{e}}^{i {\bf{k}}\cdot(x{\hat{\bf{x}}}+y{\hat{\bf{y}})}}
    \label{eqn:fft}
\end{equation}
where the complex function $q(x,y,z,t)$ can represent any physical quantity. To extract the spatial structure of the convective modes, we consider one-hour averages in time.  This gives us (with a slight notational abuse)
\begin{equation}
    q_{\mathrm{mode}}(x,y,z)= \widetilde{q}(z) {\mathrm{e}}^{i {\bf{k}}\cdot(x{\hat{\bf{x}}}+y{\hat{\bf{y}})}}
\end{equation}
where the phase of $q_{\mathrm{mode}}$ is now arbitrary. We remove this arbitrariness by multiplying $\widetilde{q(z)}$ by 
$\frac{|\widetilde{v_z}(z=-0.98~\mathrm{Mm})|}{\widetilde{v_z}(z=-0.98~\mathrm{Mm})}$.
We then have 24 realizations of one-hour averages, which we again average.
Since the simulation does not contain rotation, neither the $x$ nor $y$ direction is preferred, and so in analysing each mode we can rotate the coordinate system so that $k_y=0$. In this case the $y$ dependence of the modes is lost, and we obtain
\begin{equation}
    q(x,z)= \widetilde{q}_{k}(z) {\mathrm{e}}^{i k x}.
\end{equation}
Since we have multiple modes for each $k$, we can further reduce the realization noise by averaging over $k$. Assuming the modes change slowly with $k$, we bin in $k$ by averaging $\widetilde{q}_{k}(z)$ for those $k$ in the range $n \le k \times\left( \frac{60~\mbox{Mm}}{2\pi} \right) < (n+1)$ where $n$ is an integer. In this study we consider average modes corresponding to $n=1$ to $n=82$ and shall use $n$ to label the modes averaged in this way.

We find that the energy of the vertical motions of the eigenfunction defined here as $\sum_z{(v_z)^2}$ where the sum is over the heights indicated in Fig.~\ref{fig:radial_eigenfunctions} is almost entirely in the real part, with $\sum_z{(\Im m(v_z))^2}/\sum_z{(\Re e(v_z))^2} <0.0013$ for each of the 82 average modes we have constructed. We have here used $\Re e(...)$ and $\Im m(...)$ to denote the real and imaginary parts respectively. Similarly we find $\sum_z{(\Re e(v_x))^2}/\sum_z{(\Im m(v_x))^2} <0.00017$. Our conclusion is therefore that this method of extracting the convective modes introduces little noise, in our particular case.  

\subsection{Fourier analysis and Power spectra}
To calculate the power spectra, we use Eqn~\ref{eqn:fft}. We use two different time averages one with $24$ hour averaging where we average the simulations over the whole $24$ hour period, and one with $12$ minute averaging, where we first break the $24$ hour time series into $120$ separate $12$ minute segments. We then calculate the power spectra for each segment, and then average the 120 segments.

\section{Results}
\subsection{Mode Structure}
\begin{figure}
  \noindent
\includegraphics[scale=0.45]{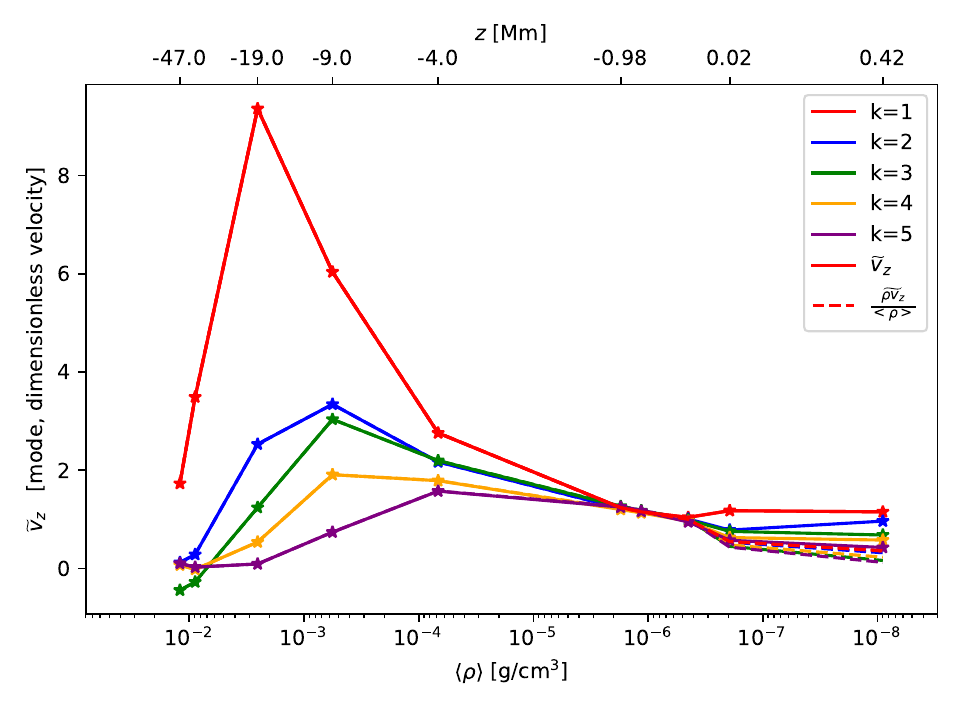}  

\includegraphics[scale=0.45]{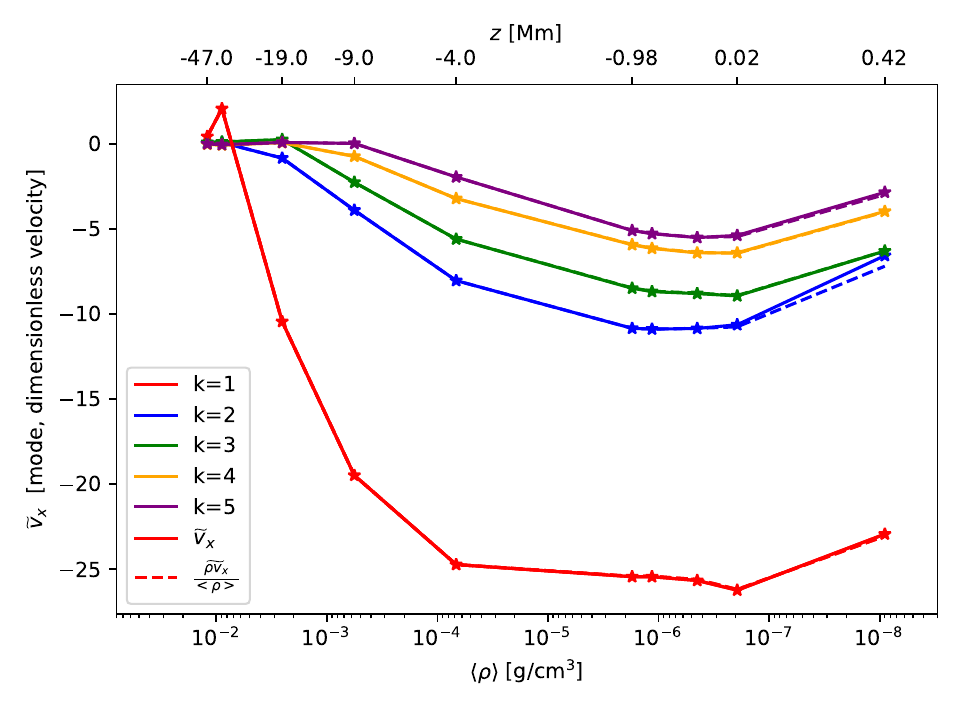}
\caption{Radial structure of convective modes with different wave numbers $k$ of the $60$~Mm~$\times 60$~Mm box. 
The spatial structure of the vertical velocity of each mode is given  $\bf{v}=\widetilde{\bf{v}}(z) \mathrm{e}^{i k x}$. The top panel shows the radial structure of the vertical velocity $\Re({\widetilde{v_z}})$, the bottom panel the horizontal velocity $\Im(\widetilde{v_x})$. 
In both panels we show the mode structure for both $\widetilde{v}$ and $\widetilde{\rho v}/\langle \rho\rangle$. The difference between the two is negligible except near and above the surface where $\widetilde{\rho}\langle v_x \rangle$ becomes substantial. The velocities of each mode were scaled
so that $\widetilde{v_z}|_{z=-0.4~\mathrm{Mm}}=1$.
\label{fig:radial_eigenfunctions}}
\end{figure}

Figure~\ref{fig:radial_eigenfunctions} shows the vertical structure of the velocity $\widetilde{v_z}$ and $\widetilde{v_x}$. Also shown are the density weighted velocities $\widetilde{\rho v_z}/\langle \rho \rangle$ and $\widetilde{\rho v_x}/\langle \rho \rangle$, where 
$\langle ... \rangle$ denotes the depth dependent mean of a quantity. In the case where the density perturbation associated with the mode is small, we would have $\widetilde{\rho v_z}/\langle \rho \rangle=\widetilde{v_z}$. This is the case except near the surface and only for the $z$ component of the velocity.

In general,
\begin{equation}
\left(\widetilde{\rho v_z}\right)_k=\sum_{n=-\infty}^\infty (\widetilde{\rho})_n (\widetilde{v_z})_{\frac{kL}{2\pi}-n}      
\end{equation}
where $L=60$~Mm, so that a large number of terms could contribute. Fortunately the numerical simulations indicate only two terms are important:
\begin{equation}
\left(\widetilde{\rho v_z}\right)_k \approx  (\widetilde{\rho})_0 (\widetilde{v_z})_k 
+(\widetilde{\rho})_k (\widetilde{v_z})_0      
\end{equation}
Noting $(\rho)_0=\langle \rho \rangle$ and 
$(v_z)_0=\langle v_z \rangle$, we can now see that we can ignore the density perturbation associated with the mode when $\frac{\widetilde{\rho}_k}{(\widetilde{v_z})_k} \ll \frac{\langle \rho\rangle }{\langle v_z \rangle}$. Where the left-hand side being a property of the modes and the right-hand side a property of the background. We find that the density perturbations become large (in the above sense) at supergranular scales.
The term $\langle v_z \rangle$ arises because although statistical equilibrium requires $\langle \rho v_z\rangle=0$, correlations between $v_z$ and $\rho$ allow $\langle v_z \rangle$ to be non-zero.

\subsection{Power spectra of vertical velocity}
\label{sec:DC}
\begin{figure}
\noindent
\includegraphics[scale=0.35]{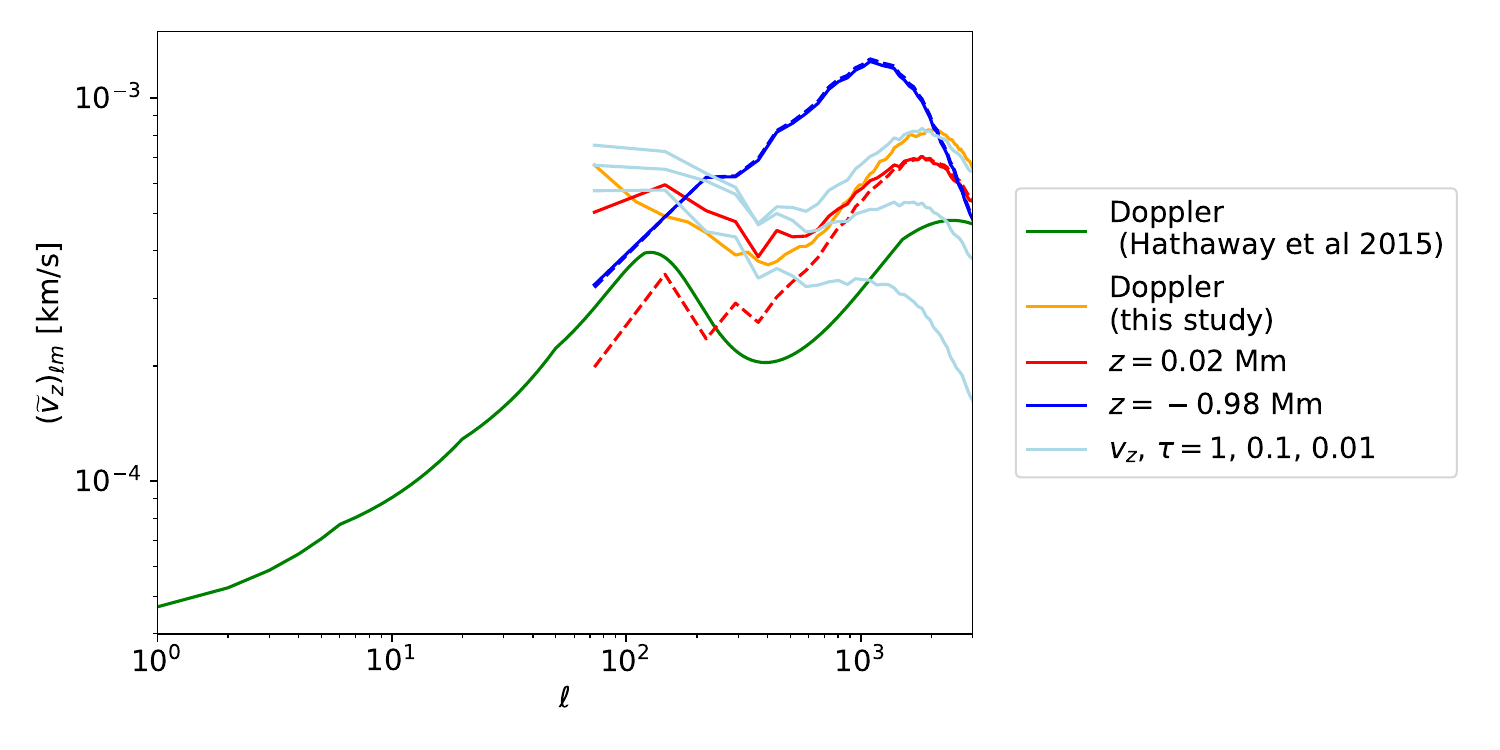}
\caption{Comparison of the power of the vertical velocity at different spherical harmonic numbers. The orange curve shows the spectra of an approximately  $60$~Mm~$\times 60$~Mm box extracted from the centre of HMI full disk Dopplergrams, the solid green curve is based on \cite{HathawayTeil2015}. The light blue curves show the vertical velocity at $\tau_{5000}\approx1$, $0.1$ and $0.01$. The red solid curve shows the spectra calculated from the vertical velocity at a constant geometric depth slightly ($20$~km) above the solar surface, the red dashed curve shows the spectra of $\widetilde{\rho v_z}/\langle \rho \rangle$. The two differ because of the density perturbation associated with the mode. The difference disappears by a depth of about $1$~Mm ( overlapping dashed and solid blue curves).
 \label{fig:vr}}
\end{figure}

The difference between $\widetilde{v_z}$ and $\widetilde{\rho v_z}/\langle \rho \rangle$ shown in Fig.~\ref{fig:radial_eigenfunctions} is based on the velocity at fixed geometric heights. We find the issue also to have an impact on the velocity determined on surfaces of constant optical depth (see Fig.~\ref{fig:vr}). 
We note that to make this figure we have to relate $\widetilde{v}_k$ to $\widetilde{v}_\ell$ since the MURaM simulations are calculated in a Cartesian box with dimensions $60$~Mm in each of the horizontal directions
rather than for the whole Sun. The discrete Fourier transform gives power at discrete wave numbers $k$. \cite{BirchProxauf2024} use
\begin{eqnarray}
 \langle v^2\rangle &=& \sum_{\bf k} \widetilde{v}^2_{\bf k}\\
 &\approx&\sum_\ell \widetilde{v}^2_{(\ell/R)} \left(\frac{\Delta \ell}{R_\odot \Delta k}\right) 
 \end{eqnarray}
where $\widetilde{v}^2_{\bf k}$ represents the power in one individual 2D Fourier mode. The wavenumber of the box is $\Delta k =2 \pi / L$, so
\begin{eqnarray}       
 \sum_{\bf k} \widetilde{v}^2_{\bf k}
 &\approx&\frac{L}{2\pi R_\odot} \sum_\ell \widetilde{v}^2_{(\ell/R)}
\end{eqnarray}
or
\begin{equation}
\widetilde{v}^2_{\ell/R_\odot} \approx \frac{L}{2\pi R_\odot} \widetilde{v}^2_{k},
\label{eqn:k2l}
\end{equation}
where the average implicitly defined here is over all modes with $|{\bf k}|\in [k,k+\Delta k)$ and the implied average over $\ell$ is over those modes with 
$\ell\in [\ell,\ell +\Delta \ell)$,
which becomes those modes with a given $\ell$ when $\Delta \ell=1$.

The error in this approximation must differ for the 
two components $\widetilde{v}_x$ and $\widetilde{v}_z$ 
of $\widetilde{\bf{v}}$. Explicitly, we will later see that (except at the surface) $\ell \widetilde{v_x}\sim  \widetilde{v_z}$, and so the error in the approximation must differ between $\widetilde{v}_x$ and $\widetilde{v}_z$. Since it is the radial transport of enthalpy which is driving the motions, we expect   
the application of Eqn.~\ref{eqn:k2l} to $\widetilde{v_z}$
is the more valid of the two, and so we use it here.
We see that the vertical velocity from the simulations on the surfaces of $\tau=1$, $0.1$ and $0.01$ resembles that of $\widetilde{v_z}$ at $z=20$~km. 
To compare with Doppler observations we used the Helioseismic and Magnetic Imager \citep[HMI,][]{Scherrer+2012,Schou+2012} full-disk magnetograms from 5 December 2019 to 14 December 2019 inclusive, using 12-minute averaged Dopplergrams per day taken at 4, 12 and 20 hours (UT). From each image we extracted the patch centred at the disk centre and measuring 
$60$~Mm by $60$~Mm in longitudinal and latitudinal directions. These patches are close enough to disk centre for the non-radial components to be negligible compared to the radial motions.

Applying the same techniques as above, we arrive at the orange curve in Fig.~\ref{fig:vr}. The observed Doppler velocity power spectrum is thus consistent with $\widetilde{v_z}$ at $20$~km at the given $\tau$ surfaces. All these spectra show distinct bumps at both granular and supergranular scales. What does not match is the spectrum of 
$\widetilde{\rho v_z}/\langle \rho \rangle$, which shows only a plateau at the supergranular scales. This behaviour is also seen in the spectrum of both $\widetilde{v_z}$ and $\widetilde{\rho v_z}$ at a depth of $0.98$~Mm. We have also plotted the spectrum from \citep{HathawayTeil2015}, who calculate the average velocity for each $(\ell,m)$ mode from the data and formulae in that paper. We note that the Doppler spectrum was calculated based on projecting the Doppler velocities from the visible side of the Sun onto a longitudinal and latitudinal mask, then taking a spherical harmonic transform. This introduces a factor of two reduction in the amplitudes which we correct for when comparing to the LCT and simulations. We calculate the average velocity for each $(\ell,m)$ mode from the data and formulae in \cite{HathawayTeil2015}. We comment that although \citeauthor{HathawayTeil2015} removed the convective mean blue shift as a function of position from the disk centre, this does not remove the impact coming through the $\widetilde{\rho}_k$. 

To qualitatively understand the effect of supergranular density perturbations on the Dopper velocity at the supergranular scale, we present a simple model consisting of three scales. Firstly we have background quantities $\rho_0$ and $v_{z,0}$, then quantities at the scale of granules 
$\rho_G$, $v_{z,G}$ and finally at the scale of supergranulation $\rho_{SG}$, $v_{z,SG}$. We introduce typical wavenumbers $k_G$ and $k_{SG}$ for the granules and supergranules respectively. Then
\begin{eqnarray}
v_z&=&v_{z,0}+v_{z,G}+v_{z,SG}\\
\rho&=&\rho_{0}+\rho_{G}+\rho_{SG}
\end{eqnarray}
Ignoring the time dependence of the density, the continuity equation gives
\begin{eqnarray}
\iint v_z \rho \hspace{1mm} \mathrm{d}x~ \mathrm{d}y=0
\end{eqnarray}
where the integration is over any horizontal surface.
Since at a constant geometrical height the density perturbation are in phase with the velocity perturbations near the surface downflows are less dense then their surroundings), we have 
\begin{eqnarray}
\rho_0 v_{z,0}+\frac{\rho_G v_{z,G}}{2} +\frac{\rho_{SG} v_{z,SG}}{2}=0
\end{eqnarray}
Because supergranulation has much weaker flows than granulation, we can obtain
\begin{eqnarray}
v_{z,0}=-\frac{\rho_G v_{z,G}}{2 \rho_0} 
\end{eqnarray}
This is conceptually related to, but not quite the same as, the Sun's convective blue shift. 
In order to estimate the Doppler shift at the supergranular scale which is not associated with a bulk mass motion, we can consider the case when
\begin{eqnarray}
(\rho v_z)_{SG}=0
\end{eqnarray}
where $(\rho v_z)_{SG}$ is the mass flux at the scale of supergranules. Now
\begin{eqnarray}
(\rho v_z)_{SG}=\sum_{k+m=SG} \rho_k v_{z,m} 
\end{eqnarray}
which turns out to be approximately
\begin{eqnarray}
(\rho v_z)_{SG}= \rho_0 v_{z,SG}+\rho_{SG} v_{z,0}
\end{eqnarray}
hence
\begin{eqnarray}
v_{z,SG}&=& -\frac{v_{z,0}}{\rho_0} \rho_{SG}\\
        &=& \frac{\rho_G v_{z,G}}{2 \rho_0}^2 \rho_{SG}
\end{eqnarray}
which gives rise to a Doppler shift even for the case where there is no mass transport at the supergranular scale.
This explains the behaviour seen in Fig.~\ref{fig:vr}.

We note that we see a similar behaviour between the $v_z$ and the disk center Doppler shift of the \ion{Fe}{I} $6173\;$\AA spectral line used by HMI. However due to the simplified grey radiative transfer used in the simulation, we do not perform a complete reproduction required to make a synthetic HMI observation.

\subsection{Power spectra of horizontal velocity}

\begin{figure}
  \noindent
\includegraphics[scale=0.35]{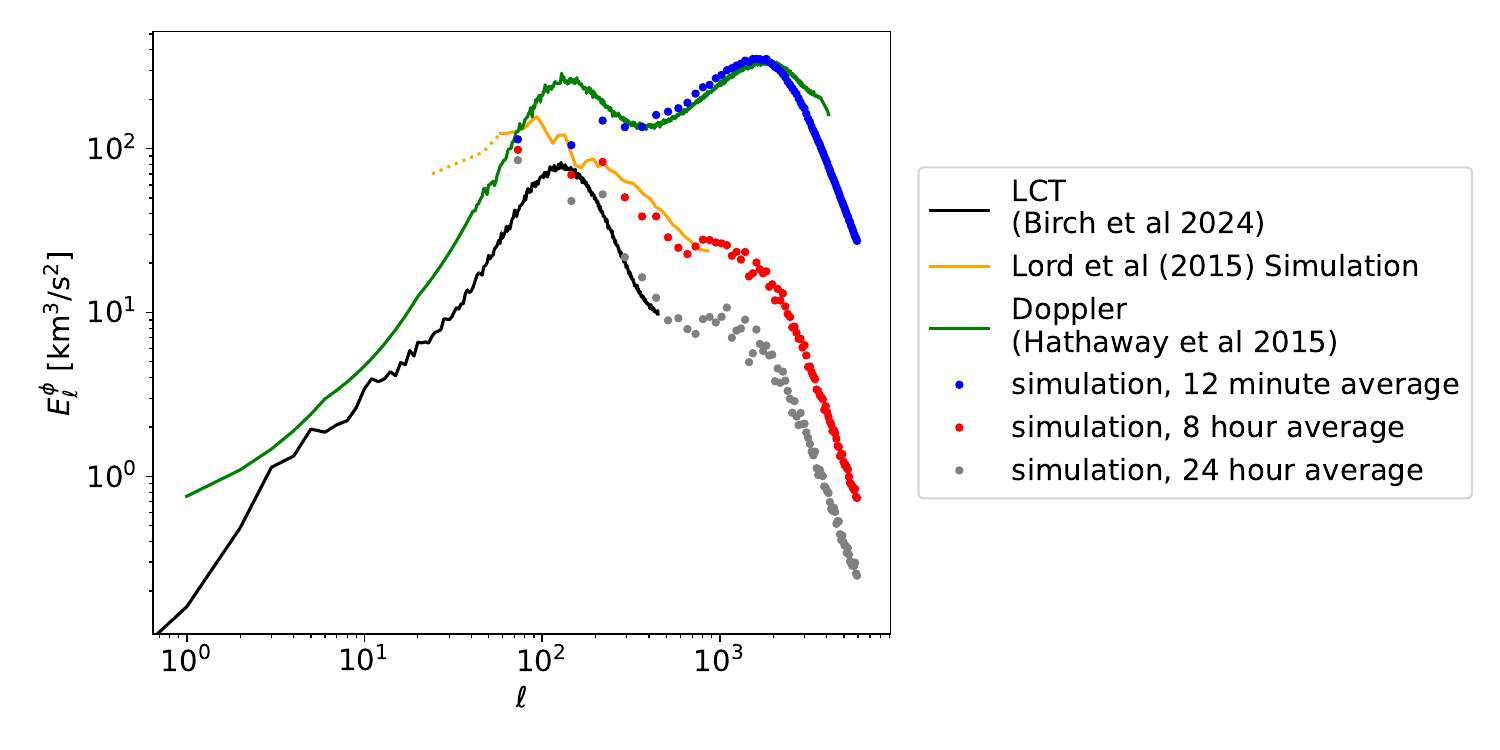}
\caption{The power of the horizontal flows as a function of wavenumber, calculated as in \cite{BirchProxauf2024}. Shown are results derived from  Doppler measurements by \cite{HathawayTeil2015} (12 minute averages), local correlation tracking by \cite{BirchProxauf2024} (27 Hour averages),
The non-magnetic simulations of 
\cite{LordCameron2014} (8 Hour LCT averages of the intensity timeseries from their MURaM  simulations), and from the current MURaM simulations at a fixed geometrical height $z=20$~km.}
\label{fig:ps_lord}
\end{figure}

The spectra of the horizontal velocity defined and calculated in the way described in \cite{BirchProxauf2024} is shown 
in Fig~\ref{fig:ps_lord} for the spectra inferred from 12 minute Doppler measurements by \cite{HathawayTeil2015}, from 27 hour averaged Local Correlation Tracking (LCT) measurements \citep{BirchProxauf2024}, and from the MURaM simulations described above, as well as the non-magnetic MURaM simulation analysed in \cite{LordCameron2014}. The 
curve from \cite{LordCameron2014} correspond to the LCT procedure being applied to the intensity images from the MURaM simulation, and thus to 8 hour averages.

\begin{figure}
  \noindent
  \includegraphics[scale=0.35]{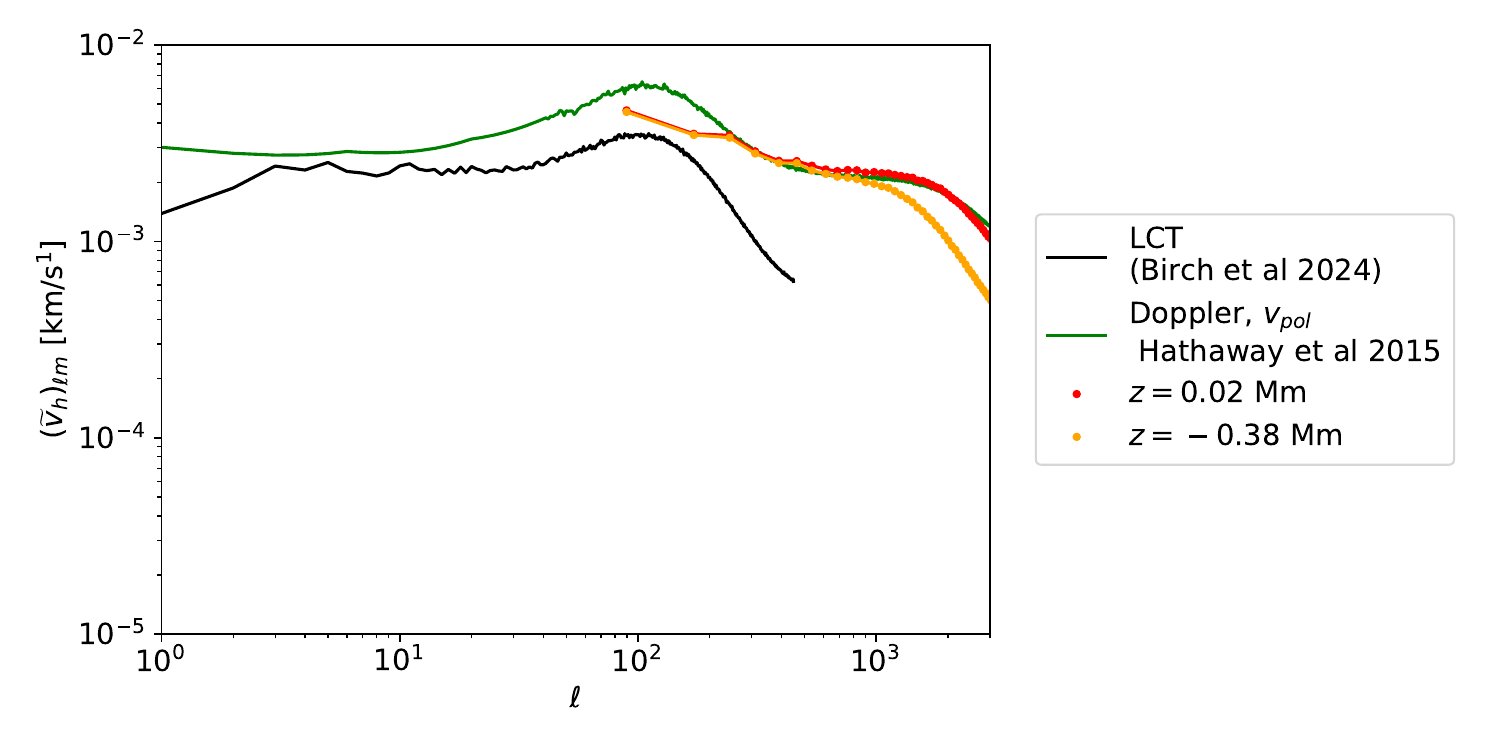}

\includegraphics[scale=0.35]{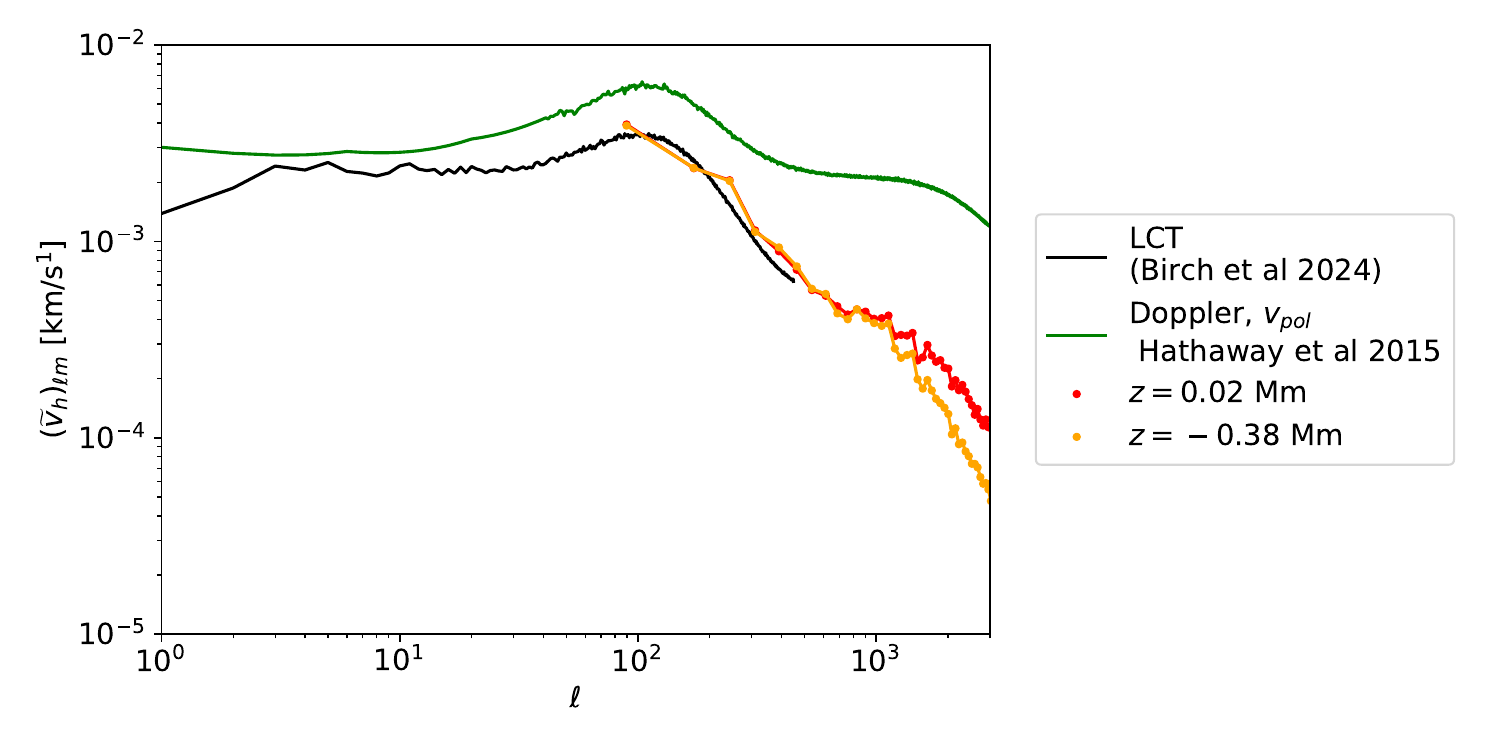}  
  \caption{Comparison of the power of the horizontal velocity at different spherical harmonic numbers. The black curve is derived from the LCT data from \cite{BirchProxauf2024}, the green curve from \cite{HathawayTeil2015}, and the blue and red curves from the MHD simulation presented here. The velocity was converted from $k$ to $\ell$ as described by Eqn.~\ref{eqn:k2l}.  The upper panel shows simulation results for a 12 minute average (corresponding to that used in Hathaway et al), the bottom panel for a 24 hour average (corresponding to that used in Birch et al.)   
 \label{fig:vx}}
\end{figure}

There are discrepancies between the new MURaM results and both the results from the Doppler inferred spectra and the LCT spectra.
The discrepancy between the spectra from the LCT and simulations can be accounted for by the way the results of the spectra is converted from $k$ to $\ell$. To see this we consider the
spectra of the horizontal velocity $\widetilde{v}_h=\sqrt{(\widetilde{v_x}^2 +\widetilde{v_y}^2)/2}$, which is shown in Fig.~\ref{fig:vx}. 
For the simulations we have calculated averages using a $k$ weighting of  $\widetilde{v_z}$, 
\begin{equation}
(\widetilde{v_h})^2_{\ell/R_\odot} = \frac{2\pi k^2 R_\odot^3}{\ell^2 L} (\widetilde{k v_h})_{k}^2.
\label{eqn:k2l_h}
\end{equation}
This treatment differs from that in \cite{BirchProxauf2024}.
For the Doppler measurements of Hathaway we use the fit and formulae in \cite{HathawayTeil2015}, again correcting for a factor of two in the different definitions of 
$\langle v^2 \rangle$.  
The spectrum by \cite{HathawayTeil2015} is mostly flat except for a bump at supergranular scales. 
The  LCT spectrum also has a bump at super granular scales, and is flat at larger scales. However at smaller scales the power falls away due to the 27 hour temporal averaging. 
The top panel shows the simulated power spectrum with a 12 minute averaging (similar to that used by \cite{HathawayTeil2015}, the bottom panel shows the spectrum when the simulations are averaged over 24 hours (similar to \cite{BirchProxauf2024}.

The match between the observations and simulations is again good, aside for the Hathaway spectrum near the supergranular bump. The bump seen in the Doppler is not seen in the velocity from the simulation at fixed height. The simulations do however match very well with the 24 hour averaged LCT measurements. That the simulations, when averaged over 24 hours, match the LCT observations, but show a discrepancy, when averaged over 12 minutes, with the velocities inferred from Doppler measurements is noteworthy.

\begin{figure}
  \noindent
  \includegraphics[scale=0.6]{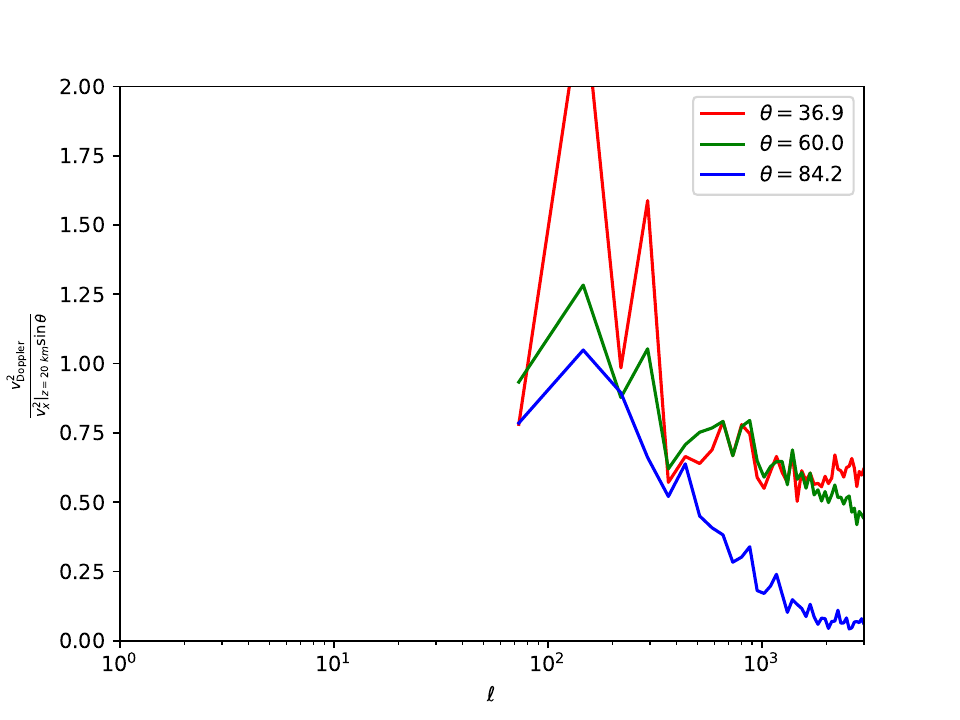}
  \caption{
  Ratio of the power spectrum calculated from simulated Doppler shifts of the FeI~6173 line, calculated from a MURaM cube at different viewing angles using the spinor code, and the respective horizontal component of the velocity in the simulation multiplied by the  sine of the viewing angle $\theta$. This analysis is based on a single snapshot from the simulation which was performed in the grey approximation.
}
\label{fig:SG_horizontal}
\end{figure}

To investigate this discrepancy we synthesized the \ion{Fe}{I} $6173\;$\AA line at three different viewing angles $\theta=84.2,60,~\mathrm{and}~36.9$, corresponding to a cosine of $\mu=0.1,0.5~\mathrm{and}~0.8$.
We compared the power spectra of the inferred Doppler velocity (determined via the centre-of-gravity method).
Because this requires the full atmosphere, we simulated it for a single snapshot.  We then computed the ratio of this Doppler velocity with that of the simulation at a constant geometric height about $20$~km above the solar surface. The result is shown in Fig.~\ref{fig:SG_horizontal}. The ratio between the horizontal flows inferred by Doppler shifts and those at a constant geometric height increases for $\ell<300$ even for modest viewing angles. 

The reason is relatively easy to understand: when viewed at an angle, the front side of granules partially hide the back side, obscuring the horizontal flows away from the observer. This reduces the power in the Doppler velocity at granular scales. Supergranules, in contrast, are relatively flat and therefore both the front side and backside of the supergranular hills are almost equally well seen. This means there is little reduction in the power in the Doppler velocity compared to the actual velocities at supergranular scales. This suggests the apparent peak at supergranular scales in the Doppler analysis is actually the result of a suppression of power in the Doppler velocities (away from disk centre) at scales smaller than supergranules.

\begin{figure}
  \noindent
  \includegraphics[scale=0.35]{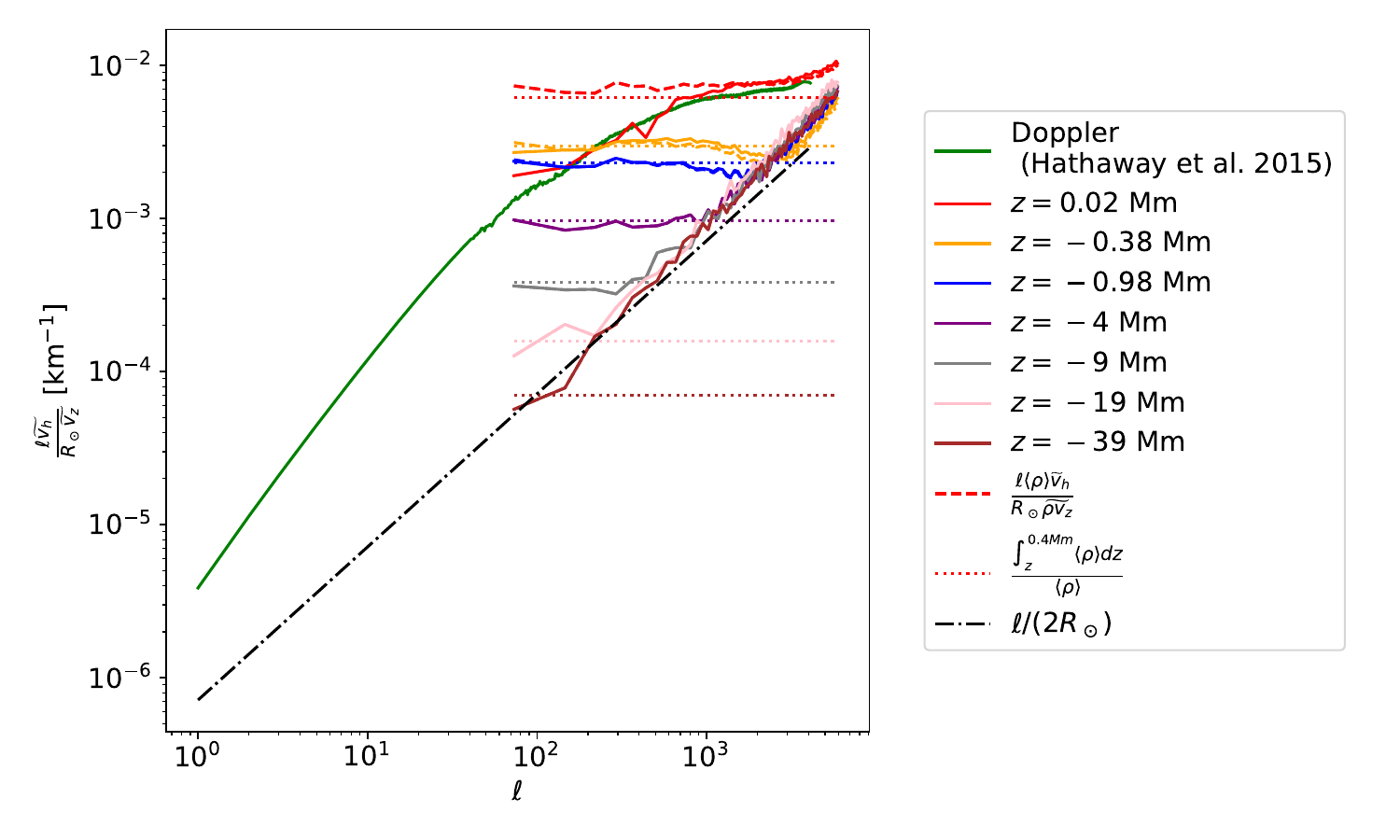}
  \caption{Ratio of the horizontal and vertical velocities, scaled by $\ell /R_\odot$ at different scales inferred from direct Doppler measurements at the surface by \cite{HathawayTeil2015}, and at different geometric depths in the simulation. The curve based on Hathaway uses only the poloidal component of the horizontal velocity.
  The simulation curves correspond to 30 minute averages to reduce the impact of acoustic oscillations.
  The shown ratio corresponds to $(H_\rho)^{-1}$
  in the regime where $\frac{\partial \ln \rho}{\partial z} v_z={k \tilde{v}_h}$.}
  \label{fig:the_solution}
\end{figure}

Figure~\ref{fig:radial_eigenfunctions} shows that for each $\ell$, there is a range of heights beneath the surface where $\widetilde{v_x}$ is almost constant. For example the $k=1\times 2\pi/60$~Mm$^{-1}$ mode, $\widetilde{v_x}$ is almost constant between $z=-4$~Mm and the surface ($z=0$). For higher $k$, the range is reduced. Following \cite{LordCameron2014},  we can understand this as in terms of a convective cell, where convective cells driven at a depth $H_\rho \ll 2 R_\odot/\ell$ extend all the way to the surface.

The fact that the horizontal velocities are almost $z$ independent in this regime is consistent with the idea that the flow is minimizing its kinetic energy density subject to the constraint that it transports a certain amount of mass above the convection cell. Based on the horizontal velocities being larger than the vertical velocities, we can approximate the total energy of the convective flows by the contribution to the kinetic energy from the horizontal flows,
\begin{equation}
E\approx\int_{Z} \rho v_h^2 \mathrm{d}z 
\end{equation}
where $z$ is the range over which $\frac{\partial v_z}{\partial z} \approx -\frac{v_z}{H_\rho}$.
Subject to the constraint that the mass flux transport at the  height where $H_\rho \ll 2 R_\odot/\ell$ is given, it is easy to see a $z$-independent $v_h$ minimizes $E$. This result is similar to the assumption in \cite{LordCameron2014} that the flow is potential above the driving depth.

Figure~\ref{fig:the_solution} shows the ratio of the velocities ($\frac{\ell v_h}{R_\odot v_z}$ ) from Doppler observations at the solar surface, and from the simulation at different heights (note we have dropped the tilde notation). We define $z=0$ as the average height at which $\tau_{500~\mathrm{nm}}=1$.
We see that the shape of the observed
ratio is reproduced by the simulated velocities just above the solar surface. The agreement disappears when we use the power spectra from $\widetilde{\rho {\bf{v}}}/\langle \rho \rangle$. This indicates that the observed drop at large scales of this ratio is due to the $\widetilde{\rho} \langle v_r \rangle$ term rather than implying that supergranules are shallow as has been previously suggested in 
\cite{November1994} and \cite{RieutordRoudier2010}.

Apart from this, we also see the two regimes studied by \cite{LordCameron2014}: the solutions cluster along the line $\left(\frac{\ell v_x}{R_\odot v_z}\right) =\frac{\ell}{2 R_\odot}$  before turning over to become flat.
The level at which it becomes flat is consistent with the idea that the 
$\widetilde{v_x}$ becomes independent of height above the driving depth of the convective mode \citep{LordCameron2014}. In particular, assuming the evolution is slow, the continuity equation gives
\begin{equation}
    \frac{\partial {\rho v_z}}{\partial z}+\frac{\partial {\rho v_x}}{\partial x} =0
\end{equation}
which for a single mode gives,
\begin{equation}
    \frac{\mathrm{d} \widetilde{\rho v_z}}{\mathrm{d} z}+ \frac{i\ell}{R_\odot} \widetilde{\rho v_x} =0
\end{equation}
exploiting the fact that $\widetilde{\rho v_x}=\langle \rho \rangle \widetilde{v_x}$, we have
\begin{equation}
    \frac{\mathrm{d} \widetilde{\rho v_z}}{\mathrm{d} z}= -\frac{i\ell}{R_\odot} \widetilde{v_x} \langle \rho \rangle
\end{equation}
which we can integrate to obtain
\begin{equation}
    \widetilde{\rho v_z}= -\frac{i\ell}{R_\odot} \widetilde{v_x} \int_z^\infty\langle \rho \rangle dz
\end{equation}
so
\begin{equation}
    \frac{i \ell \widetilde{v_x}}{R_\odot\widetilde{\rho v_z}}= \frac{1}{\int_z^\infty\langle \rho \rangle dz}
\end{equation}
which trivially gives
\begin{equation}
    \frac{i \ell \langle \rho \rangle \widetilde{v_x}}{R_\odot\widetilde{\rho v_z}}= \frac{\langle \rho \rangle}{\int_z^\infty\langle \rho \rangle dz}
\label{eqn:lord}
\end{equation}

\begin{figure}
  \includegraphics[scale=0.36]{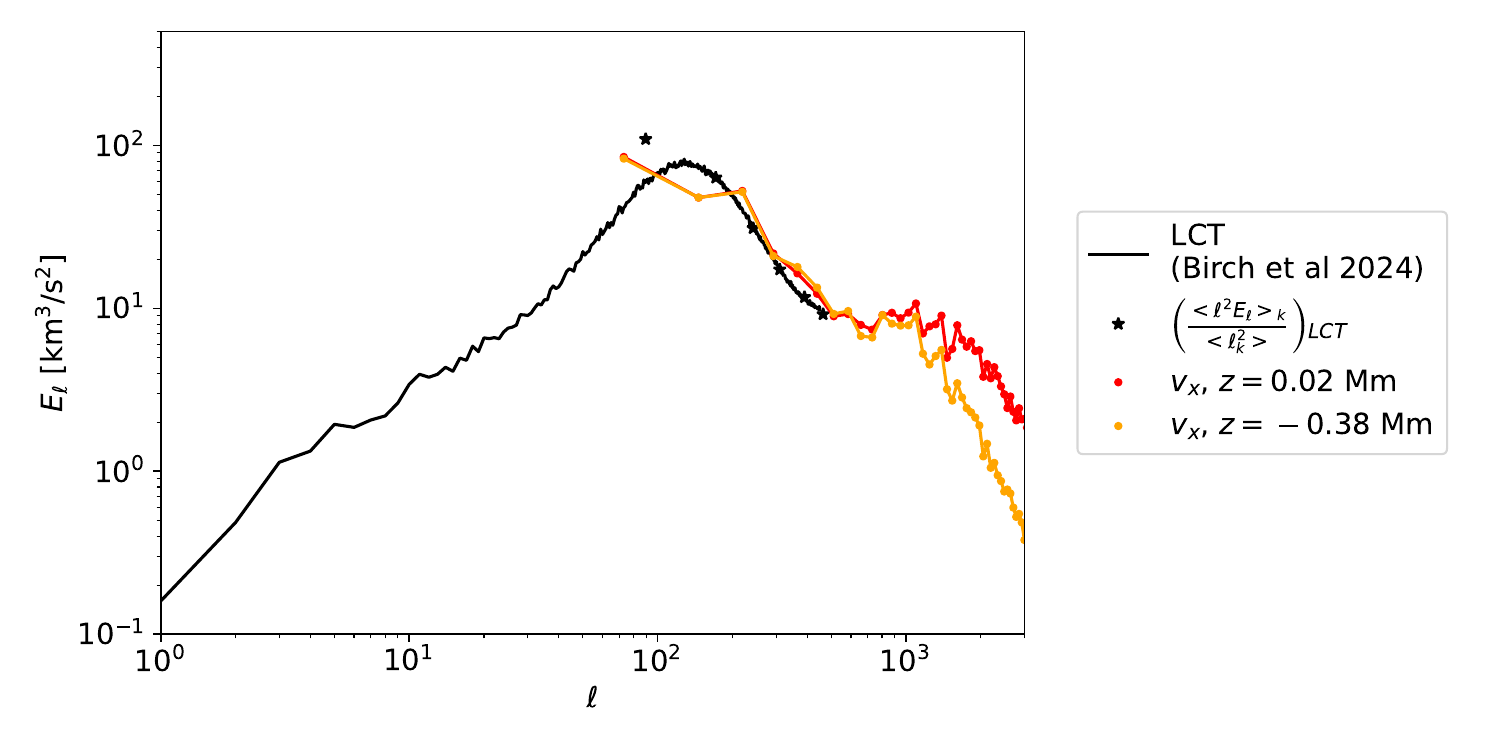}
  \includegraphics[scale=0.36]{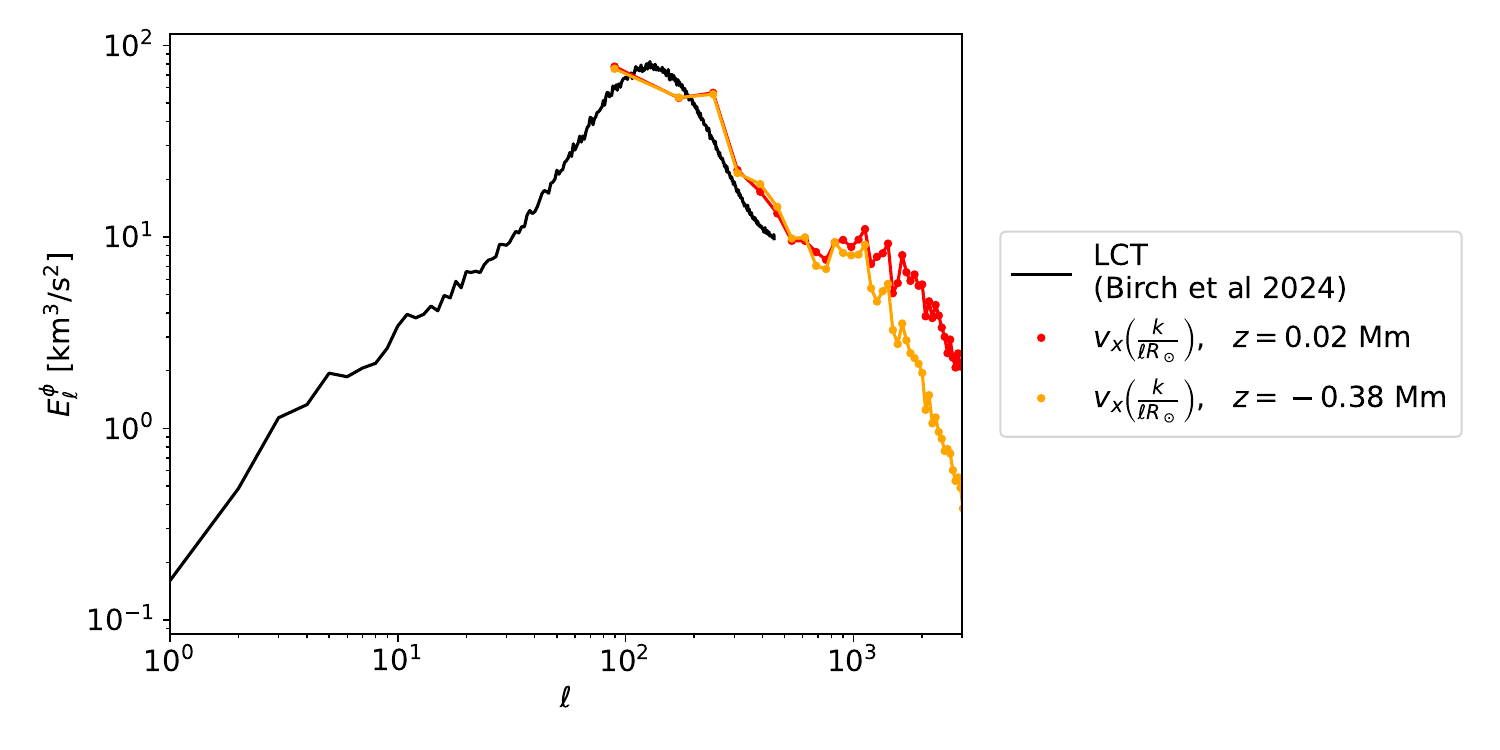}
  \caption{Power spectrum of the horizontal velocities in the format of \cite{BirchProxauf2024}. The top panel shows the results using the routines provided by \citeauthor{BirchProxauf2024}, the lower panel shows the similar results obtained using the $k-$weighted averages.
  \label{fig:arron}}
\end{figure}

The expression on the right-hand side of Eqn~\ref{eqn:lord} can be considered a scale height, and the result is an extension of the result in \cite{LordCameron2014}. Fig.~\ref{fig:the_solution} shows this result is applicable all the way to the surface for scales larger than $2R_\odot/\ell$. Finally we note that we can put Fig.~\ref{fig:vx} into the form used in \cite{BirchProxauf2024} which is presented in Fig.~\ref{fig:arron}. We have done so for $E^\phi$ based both on the spectra of $\widetilde{v}_h$ and the 
$k$-weighted average $\widetilde{k v}_h/\ell$.
The simulations nicely reproduce the observed power spectrum for all the simulated scales, that is for scales smaller than or equal to $60$~Mm. 

\subsection{Depth dependence of $(v_z)_{\ell m}$. }
\begin{figure}
  \noindent
\includegraphics[scale=0.35]{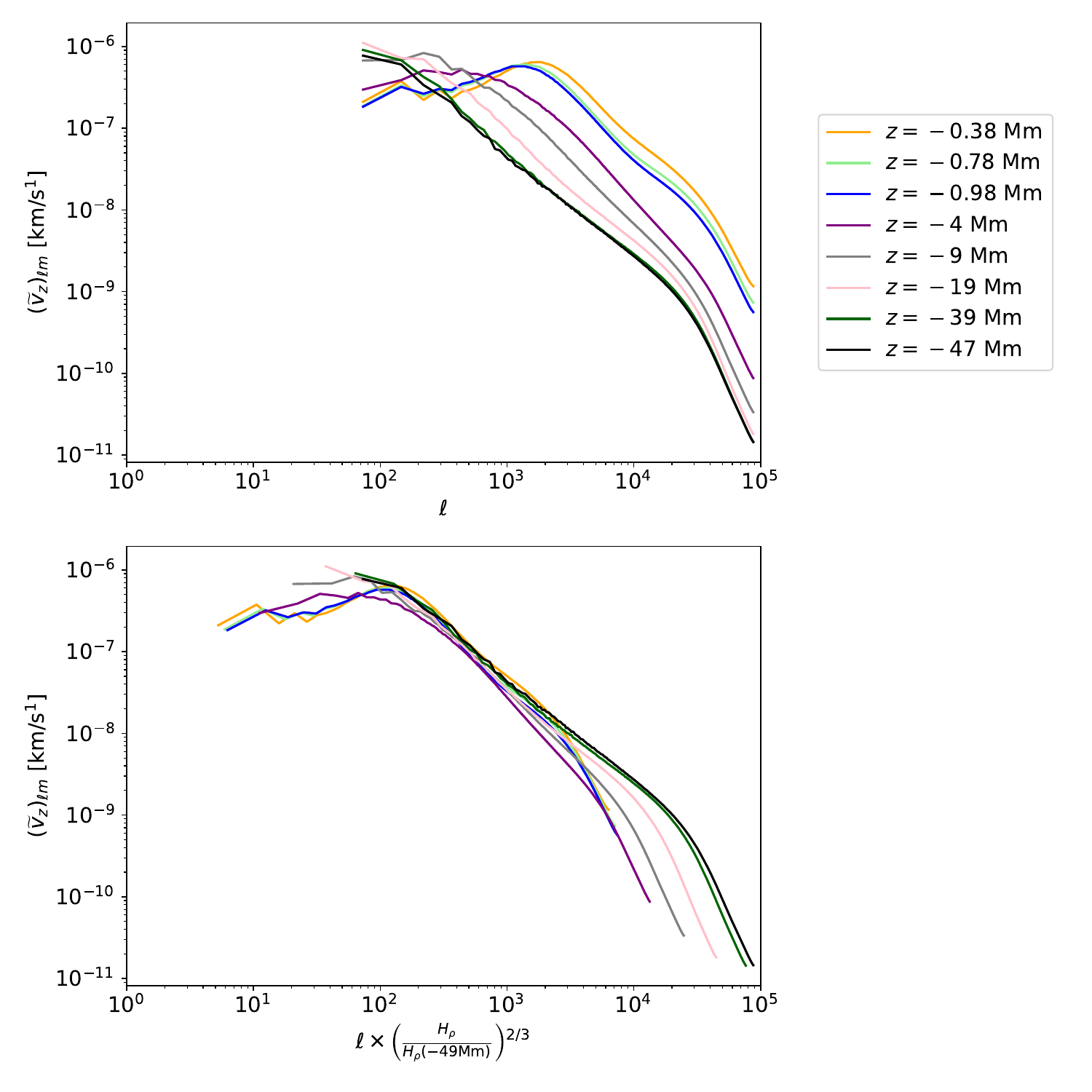}
  \caption{Vertical velocities, $v_z$, at different depths. The top panel is in the same format as Fig.~\ref{fig:vx}. The bottom panel shows the result when we plot against $\ell \times \left( \frac{H_\rho}{H_\rho\mid_{z=-49\mathrm{ Mm}}} \right)^{\frac{2}{3}}$  }
 \label{fig:vz}
\end{figure}

The radial heat flux in the convection zone is carried by radial motions, corresponding to $v_z$ in our system. The top panel of Fig.~\ref{fig:vz} shows $v_z$ as a function of $\ell$ for different depths. These collapse onto a single curve when plotted against $\ell \times H_\rho^{\frac{2}{3}}$, where $H_\rho$ is the density scale height calculated here in the usual way, $\langle \rho \rangle/\left(\frac{d \langle \rho \rangle}{dz} \right)$. The result indicates that the horizontal forcing scale of these magnetohydrodynamic simulations differs from that reported for the hydrodynamic simulations of \cite{LordCameron2014}, where a scaling with $H_P$ was found. We hence agree with \cite{HottaKusano2021} that the small-scale dynamo has strong implications in the context of the convective conundrum.
We will now show the amplitude of the vertical velocity in the simulations is consistent with mixing length (where the vertical scale of the convective mixing is proportional to the pressure scale height) with the scaling with $H_\rho^{2/3}$ only applying to the horizontal scaling.

\subsection{Mixing length prediction of the depth dependence of $(v_z)_{lm}$.} In this section we show that the vertical velocity amplitude is consistent with the assumption that the length-scale in the radial direction, $L$, scales with $H_P$ whilst the length scale in the horizontal direction scales with $H_\rho^{2/3}$. We will do this by considering both the scaling of the amplitude of the vertical velocity and the horizontal scaling associated with it. As seen in Fig.~\ref{fig:vz}, the horizontal scale of the forcing scales like $H_\rho^{2/3}$. Explicitly
\begin{eqnarray}
\left({v_z}\right)_{lm}(z)=v(z) f\left(\ell H_\rho^{2/3}\right)
\end{eqnarray}
where $v_z$ captures the $z$ dependence of the amplitude of the velocity and $f$ captures the dependence of the forcing scale.

\begin{figure}
  \noindent
  \includegraphics[scale=0.35]{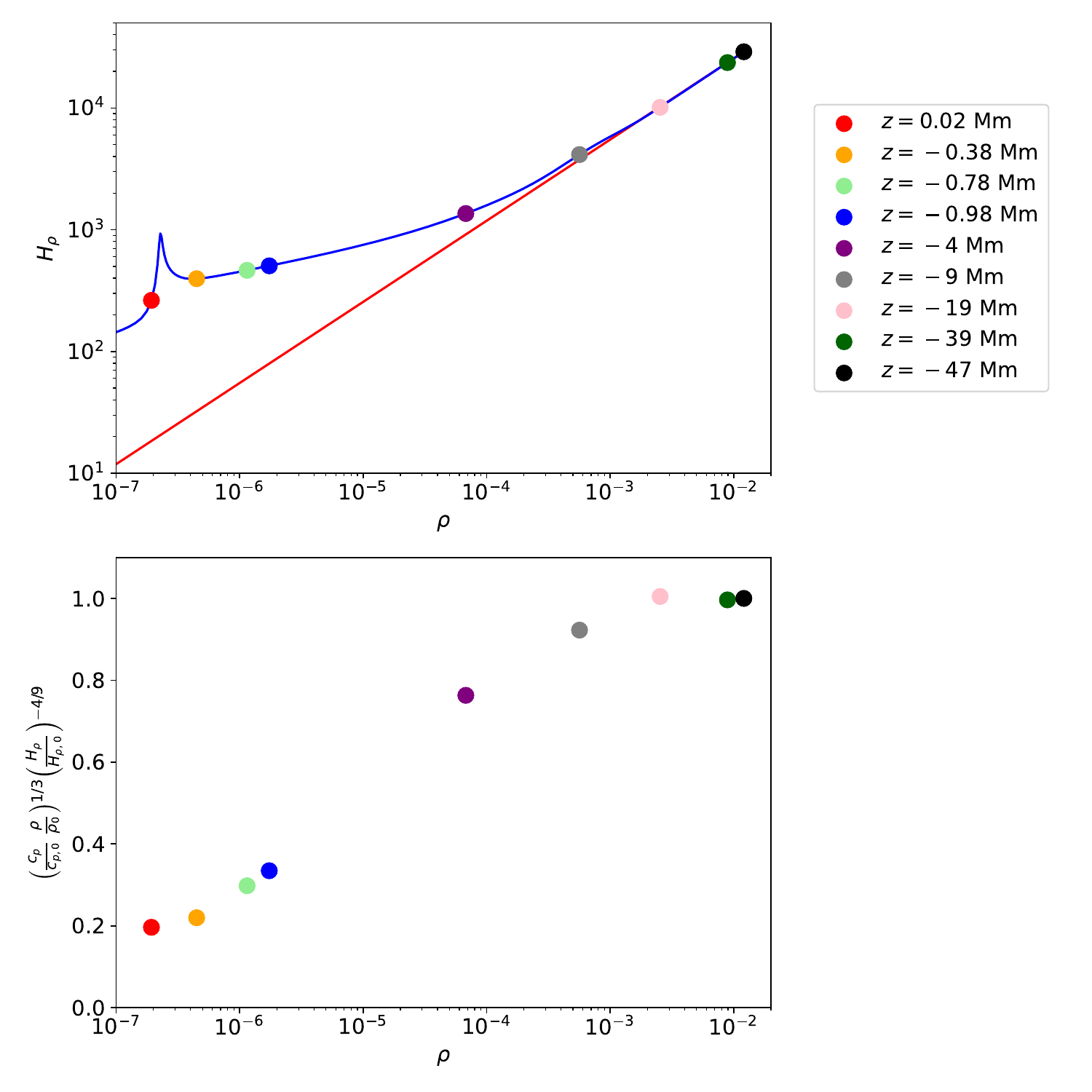}
  \caption{The top panel shows the density scale height, $H_\rho$, as a function of density, $\rho$ in blue. The red curve shows the result for a polytrope with index $1.5$. The coloured dots show the values corresponding to the different depths shown in the lower panel.  The bottom panel shows the scaling expected from mixing length  $\left(\frac{C_p}{C_{p,0}}
  \frac{\rho}{\rho_{0}}\right)^{1/3}
  \frac{H_{\rho}}{H_{\rho,0}}^{-4/9}$, where the subscript~0 indicates the values at a depth of 49~Mm. That the values are close to 1 for depths below 4~Mm explains why the curves in Fig~\ref{fig:vz} did not need to be shifted in amplitude to align.
  }
\label{fig:MLT}
\end{figure}
Away from the surface, where hydrogen is fully ionized, the expectation from mixing length theory is that the flux carried by convection per mode is
\begin{eqnarray}
F_{\ell m} \sim \frac{\mu}{\alpha} c_p \rho (v_z)_{\ell m}^3.
\end{eqnarray}
where $\alpha$ is the mixing length parameter and $\mu$ is the mean molecular weight. 
Integrating over $\ell$ and $m$ yields
\begin{eqnarray}
F &\sim& \int_\ell \frac{\mu}{\alpha} c_p \rho (v_z)_{\ell m}^3 (2\ell+1) ~\mathrm{d} \ell \\
 &\sim& \int_\ell \frac{\mu}{\alpha} c_p \rho v(z)^3 f\left(\ell H_\rho^{2/3}\right)^3 (2\ell+1) ~\mathrm{d} \ell 
\end{eqnarray} 
hich for large $\ell$ is approximately
\begin{eqnarray}
 F&\sim& \int_\ell \frac{\mu}{\alpha}c_p \rho v(z)^3 f\left(\ell H_\rho^{2/3}\right)^3 (2\ell) ~\mathrm{d} \ell 
\end{eqnarray}
substituting $\ell'=\ell H_\rho^{2/3}$
gives
\begin{eqnarray}
 F&\sim&  \frac{\mu}{\alpha} c_p \rho H_\rho^{-4/3} v(z)^3
 \int_\ell  f\left(\ell'\right)^3 
 (2\ell') ~\mathrm{d} \ell' 
\end{eqnarray}
Beneath the surface, the flux carried by convection is independent of height, so that away from the photosphere, 
\begin{eqnarray}
v(z)\sim \left(\frac{\mu}{\alpha}\right)^{-1/3} c_p^{-1/3} H_\rho^{4/9} \rho^{-1/3}
\end{eqnarray}
Both $\alpha$ and $\mu$ are expected to be constant away from the boundary, and the resulting scaling of the velocity is shown in the bottom panel of Fig.~\ref{fig:MLT}, which together with Fig.~\ref{fig:vz}
indicates that while the horizontal scale scales like $H_\rho^{2/3}$, away from the surface, the vertical scale scales like $L\sim H_P$ as per mixing length theory.

\section{Conclusions}
The MURaM simulations reproduce the observed spectra of horizontal velocities at scales of 60~Mm or less based on both 27~hour LCT measurements \citep[from][]{BirchProxauf2024}  and 12 minute averaged Doppler observations \citep{HathawayTeil2015}. We also reproduce the radial Doppler velocity measurements 
\citep{HathawayTeil2015} except for the supergranular peak.
For horizontal flows, we have provided a potential explanation for this difference in terms of a reduction of the power at scales below supergranulation associated with viewing the corrugated granulation scale structures from an angle.

For radial (vertical) flows there is also an excess of power at supergranular scales at the surface. We have shown that this excess in the radial velocities is primarily due to density perturbations associated with the convective modes, in combination with a non-zero horizontally averaged radial flow $\langle v_z \rangle$.

In terms of the ratio of the vertical and horizontal flows,
similar to \cite{LordCameron2014}, we found that the continuity equation reveals two regimes, one where the density scale height (here defined as $\frac{\int_z^\infty \langle \rho \rangle dz}{\langle \rho \rangle}$) 
is unimportant and one in which it dominates. The simple structure of the modes suggested by \citeauthor{LordCameron2014} is now seen to extend to above the photosphere. 

We comment that at the scales investigated in this work we have not seen the onset of the high amplitude modes which lie at the heart of the convective conundrum, which might set in at $\ell$ below those covered by the simulation. We have also found that the horizontal forcing at different depths scales like $H_\rho^{2/3}$. 

\begin{acknowledgements}
Both authors contributed equally.
We thank the anonymous referee for valuable comments and suggestions.
RHC acknowledges support from the European Research Council (ERC) under the European Union’s Horizon 2020 research and innovation programme (grant agreement No 810218 WHOLESUN). We appreciate the computing resources provided by the HPC systems Raven and Viper at the Max Planck Computing and Data Facility. 
D.P. was supported by the Deutsches Zentrum für Luft und Raumfahrt (DLR; German Aerospace Center) by grant DLR-
FKZ 50OU2201. RHC acknowledges useful discussions with Aaron Birch and Jesper Schou. The LCT results were kindly provided by A.~Birch, the Doppler results by D.~Hathaway. 
\end{acknowledgements}

\bibliographystyle{aa}
\bibliography{cameron}

\begin{appendix}

\section{Simulation Stability}\label{app:stability}

\begin{figure}[hb]
  \noindent
  \includegraphics[scale=1.0]{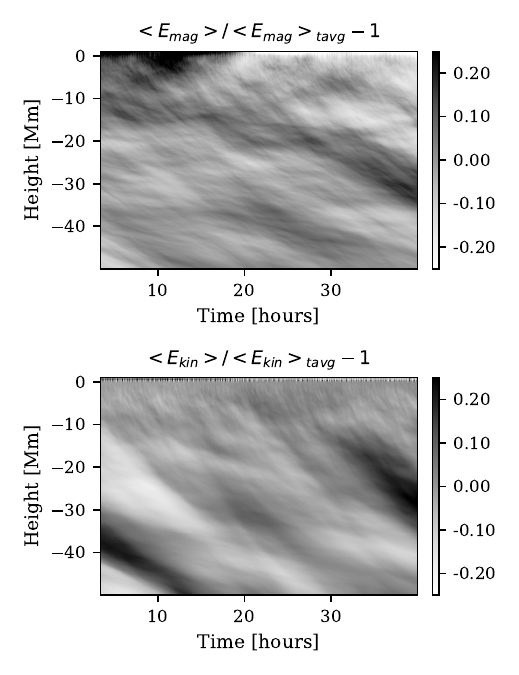}
  \caption{Time series of the horizontally averaged energy densities from the simulation normalized by the temporarily averaged value}. The top panel shows the magnetic energy, and the bottom the kinetic energy.
 \label{fig:havg_Es}
\end{figure}

\begin{figure}
  \noindent
  \includegraphics[scale=1.0]{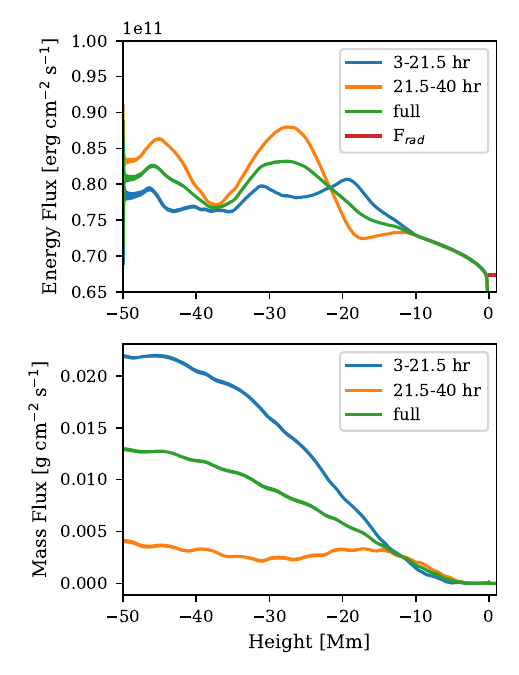}
  \caption{Horizontally and temporally averaged fluxes from the MHD simulation. The top panel shows the sum of the kinetic, magnetic and enthalpy fluxes, the bottom panel shows the mass flux. The plot shows the fluxes averaged over the full simulation (green), and also averaged only the first half (blue), and the second half (orange) of 40 hours of simulation time. The average luminosity of the box is also shown in the top panel in red.
 \label{fig:havg_fluxes}}
\end{figure}

In this section we show that the simulations are close enough to the relaxed state for the analysis performed in this work. In order to better display the evolution, the time-series shown runs 12 hours longer than the window analysed in the paper. The kinetic and magnetic energies , normalized by their time averaged value, are seen in Fig. \ref{fig:havg_Es}. No major growth in magnetic or kinetic energy is seen. The time taken for material to descend from the top to the bottom of the box is around 30 hours. In Fig. \ref{fig:havg_fluxes} we show the horizontally and temporally averaged energy flux and mass flux. We subtract the $k=0$ box mode from the kinetic and enthalpy fluxes. The total output radiative flux of the simulation is also shown. The net energy flux in the convection zone is around $12\%$ higher than the output radiative flux. This is likely caused by averaging only a single  turnover time, in combination with the small upwards mass flux that is occurring as the simulation runs. The simulation is close enough to equilibrium to not impact the conclusions of the paper.

\begin{figure}
  \noindent
  \includegraphics[scale=0.35]{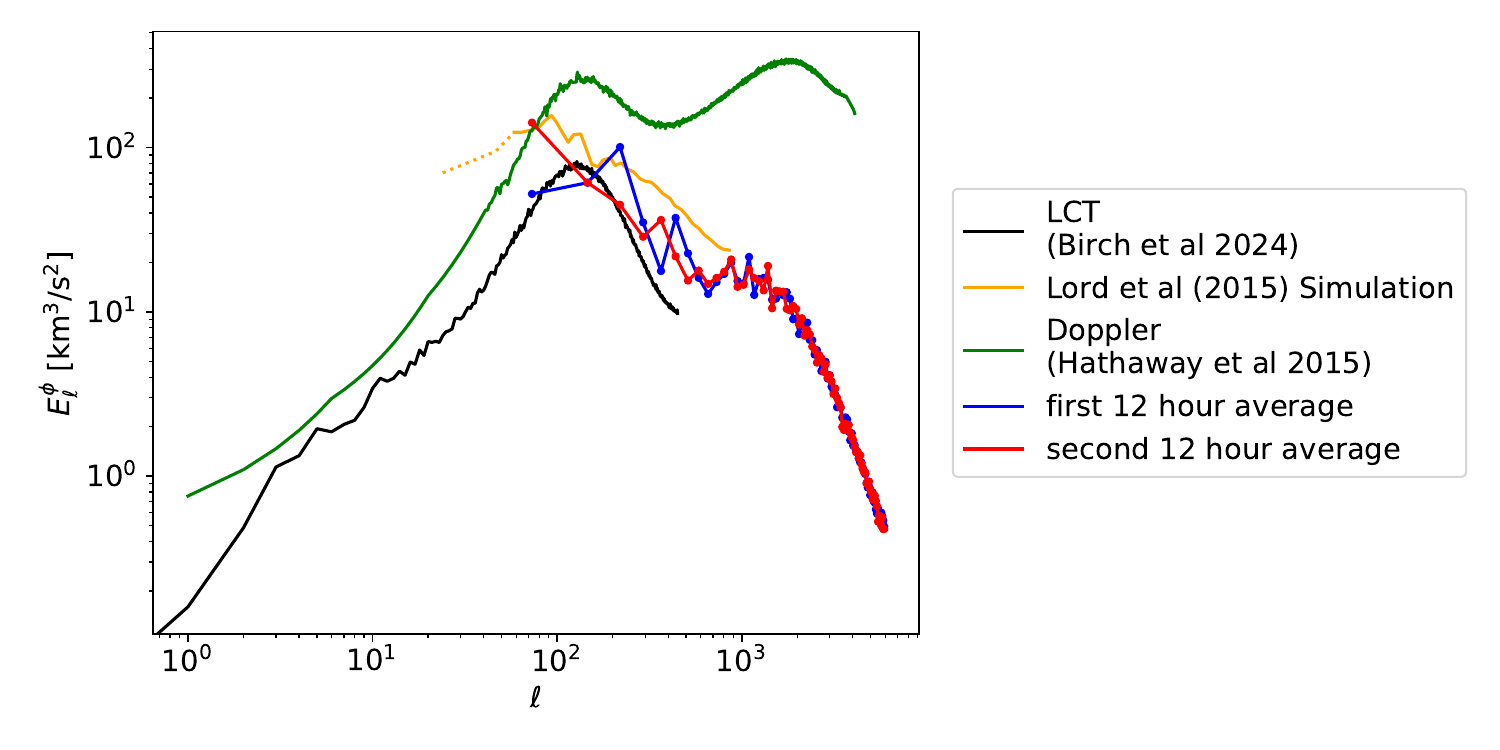}   \includegraphics[scale=0.35]{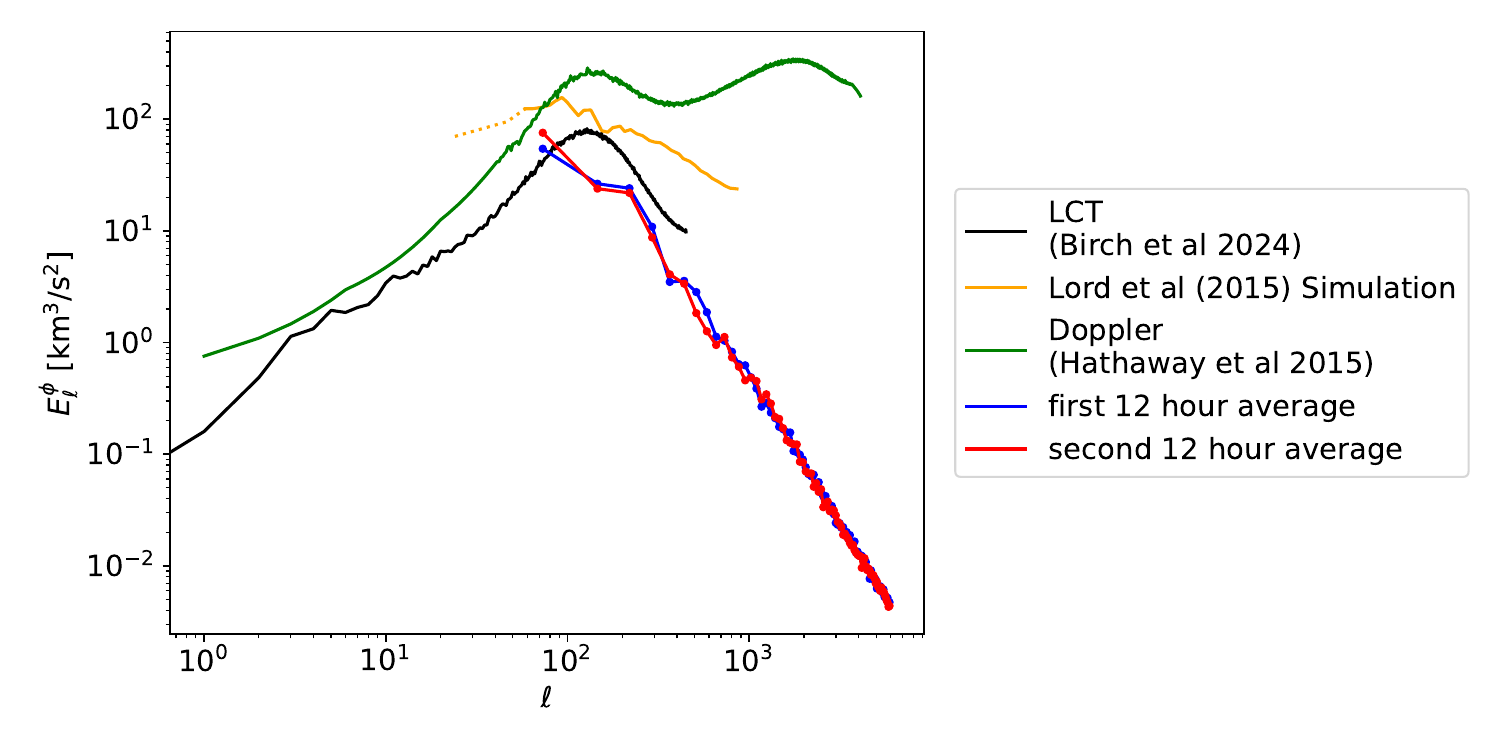}
\includegraphics[scale=0.35]{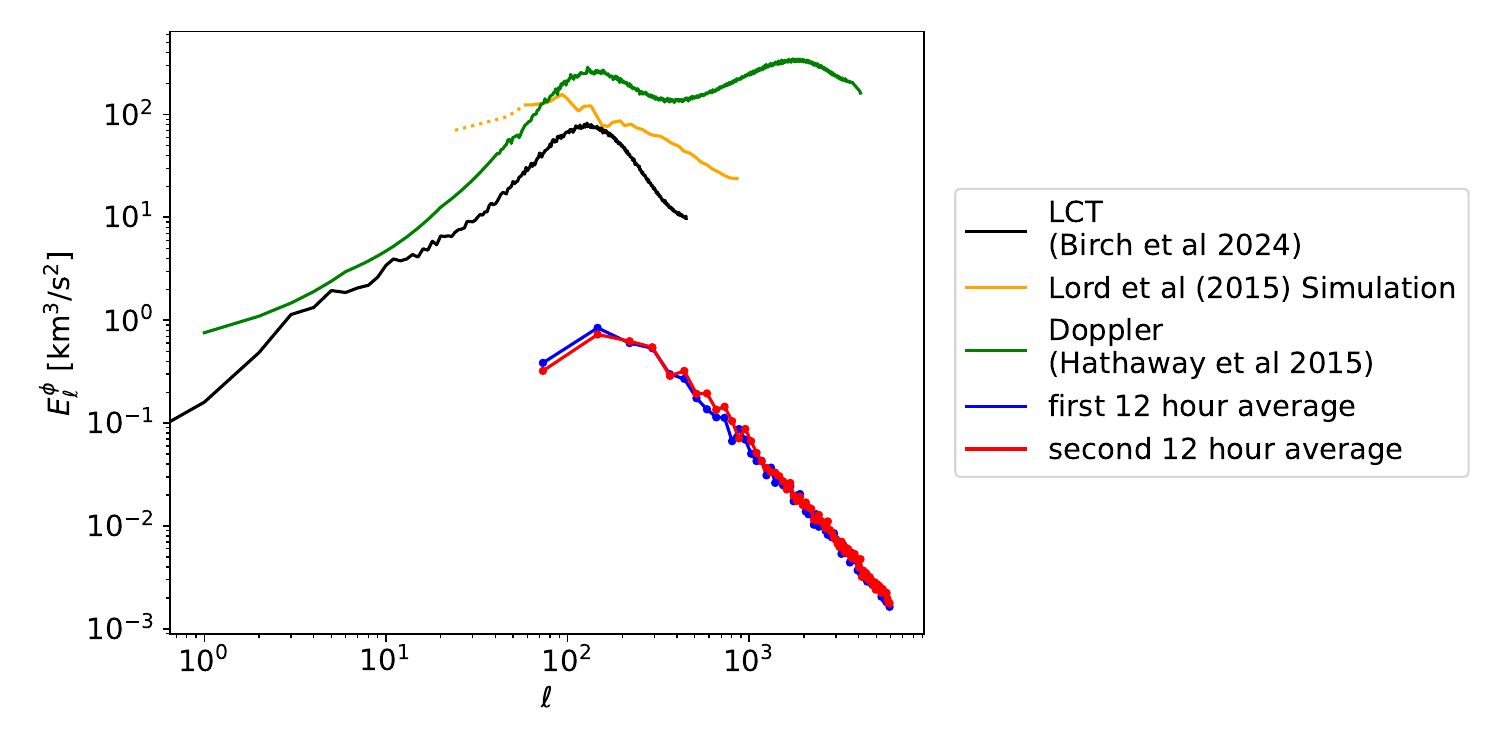}
  \caption{The same analysis as in the main text, applied to the first and second 12 hours (with a 12 hour averaging). In this case the curves are expected to lie between the black (24 hour average) and green (12 minute average) curves, obviously closer to the 24 hour average. The difference between these two curves can be interpreted as noise, with error bars of the two 12 hour averages being $\sqrt{2}$ greater than that of the 24 hour average. The values from the simulation  in the top panel are taken from a height of $z=20$~km (above the average height of the $\tau_{Ross}=1$ surface, for the middle panel from $z=-9$~Mm, and for the bottom panel from $z=-49$~Mm.
  Some contribution from evolution at long wavelengths cannot be excluded. }
 \label{fig:two_times}
 
\end{figure}

In order to have an estimate of the realization noise in the spectra (and possible longer term trends) we analysed the 24 hour timeseries discussed in the main section in two chunks, each of 12 hours. The velocities were averaged for each of the 12 hour periods and the power spectra of the average flow over both periods is shown in Fig.~\ref{fig:two_times} at 3 different heights in the box. We comment that the noise on the 24 hour averages is expected to be a factor of $\sqrt{2}$ smaller than on the 12 hour segments, and that the noise appears to be highest near the surface. We additionally comment that modes with similar $\ell$ transport energy over similar depths, and hence affect the driving of each other. This means that the realization noise can be coupled across $\ell$. 

\end{appendix}
\end{document}